\documentclass[12pt,a4paper]{article}
\usepackage{dfttrob}
\dfttnum{DFTT 36/98\\FISIST/12-98/CENTRA\\October 15, 1998}
\usepackage[standard]{refstyle}
\def\Tevatron{\textsc{Tevatron}}

\def\ee{$e^+e^-$}               % e+ e- (annihilations...)

\newcommand{\eref}[1]{(\ref{#1})}
\def\nbar{\bar n}            % NB parameters: must be used in math mode
\def\Nbar{\bar N}            %   only, or it won't work!
\def\nc{{\bar n_c}}          % extra {} ensure subscriting it is ok.

\def\pt{p\kern -.2pt\lower 4pt\hbox{\tiny T}}    %works?

\def\p0{P_0(\Delta y)}

\def\NF{\mathcal{N}_{\kern -1.9pt f}}
\def\NC{\mathcal{N}_{\kern -1.7pt c}}

\usepackage{amsmath}
\usepackage{amssymb}
\usepackage{epsfig}

\def\ksoft{k_{\text{soft}}}
\def\nsoft{\nbar_{\text{soft}}}

\def\ksemi{k_{\text{semi-hard}}}
\def\nsemi{\nbar_{\text{semi-hard}}}

\def\asoft{\alpha_{\text{soft}}}
\def\ktot{k_{\text{total}}}
\def\ntot{\nbar_{\text{total}}}

\def\rs{\sqrt{s}}

\begin{document}

\title{Possible scenarios for soft and semi-hard components  structure
            in central hadron-hadron collisions in the TeV region}
\author{A. Giovannini\\
 \it Dipartimento di Fisica Teorica and I.N.F.N. -- sezione di Torino\\
 \it via P. Giuria 1, 10125 Torino, Italy\\[3mm]
 R. Ugoccioni\\
 \it CENTRA and Departamento de F{\'\i}sica (I.S.T.),\\ 
 \it Av. Rovisco Pais, 1096 Lisboa codex, Portugal}
\maketitle

\begin{abstract}
Possible scenarios in $hh$ collisions in the TeV region are
discussed in full phase space.
It is shown that at such high energies one should expect strong KNO
scaling violation and a $\ln(\rs)$ increase of the average charged
multiplicity of the semi-hard component, resulting in a huge mini-jet
production. 
\end{abstract}

\section*{Dedication}

This paper is dedicated to Peter Carruthers.

One of the authors (A.G.) had a long correspondence with him on
the name of the distribution to the success of which in multi-particle
dynamics Peter contributed so much, and 
which is at the basis of the present work.
It can be found indeed in the literature under different names:
in particular, the name ``negative binomial'' ---currently used in
our field--- seemed too anonymous.
The agreement with Peter was to start to call it ``Pascal
distribution'', in honour of Blaise Pascal,
one of the scientists who used it first.
The passing of Peter did not allow to solidify this proposal,
which is now brought forward in his memory.

\section{Introduction}\label{sec:intro}
  
The study of final particles multiplicity distributions (MDs) and related
correlations structure in the new foreseen energy domain in the TeV region 
in hadron-hadron
($hh$) collisions is a challenging problem for multiparticle dynamics.
Here the production of events with  a huge number of final particles is
indeed the most spectacular and fascinating 
although not yet fully understood  
phenomenon. The new fact is the occurrence of high parton 
density islands in  regions  where QCD parton evolution equations cannot 
be applied, and long range correlations among produced particles are expected
to be quite large.    

A sound theory  of  strong interactions cannot avoid   to  describe  such 
complex high energy many-body system; at the same time to approach  this
problem is ancillary to the understanding of even more  complex strongly
interacting systems like proton-nucleus and heavy ions  collisions.
It should be reminded and stressed again and again that it  is still 
a problem for QCD  (in $hh$ collisions more than in others collisions 
due to the 
mentioned complexity of the reaction) how to extend the theory  from the  
perturbative to the non-perturbative sector.
Hadronization mechanism and more specifically how  to calculate from 
first QCD principles multiplicity distributions  and correlation structure 
of final particles states (the true observables) 
are here unanswered questions.
In this region where standard QCD has shy predictions for a complex system 
like hadron-hadron scattering at very high  c.m.\ energies one can rely only 
on models based  on empirical observations on multiplicity distributions
behaviour and related normalized factorial and cumulant moments both in full 
phase space, and in limited sectors of rapidity and transverse variables.
Thus
the first step of our programme is to examine critically what one learns on
$hh$  collisions in full phase space  from previous 
experimental and theoretical work  starting from accelerator region
results (in a subsequent paper we will extend our study to rapidity
intervals). 
Important hints are  expected also to come  from  other reactions like 
\ee\  annihilation and  deep inelastic scattering, where the simplicity of
the projectile and/or of the reaction itself allowed already to  isolate  
very interesting properties of the most elementary substructures (jets of
given flavor \cite{hqlett:3}) at work in  the interaction region.

The phenomenological work  done in this area is impressive.

Many regularities (like the Pascal, or negative binomial,
multiplicity distribution, Pa(NB)MD,\footnote{from now on the
abbreviation `Pa(NB)MD' will be used to mean: the Pascal (also
known as negative binomial) multiplicity distribution.}  
behaviour and KNO scaling) have been analyzed in 
all classes of collisions since 1972: this effort reminds us of the 
precious  and humble job done  by the spectroscopists generation 
on atomic spectra  in the  pre-quantum  mechanics era.
However the general agreement is only on few statements.
Nobody is questioning that particles are produced in jets, the most important
QCD prediction in the field.
In addition, observed  events are considered either ``soft'' (events without
mini-jets) or ``semi-hard'' (events with mini-jets), and the relevance of 
the  latter is expected to  increase with  c.m.\ 
energy.\footnote{Here we rest with the standard definition of mini-jets
as proposed by the UA1 Collaboration: groups of particles
having total transverse energy larger than 5 GeV.}

Limiting to central collisions only 
one can say that particles are produced not independently one from the other 
(Feynman scaling violation). Two-particle correlations are   probably 
dominant at lower c.m.\ energies; this fact  favors  here hierarchical 
correlation structure and supports the "old" Pa(NB)MD regularity: 
a "single" Pa(NB)MD 
provides indeed   a  satisfactory description of final 
particles MDs 
in the accelerator region up  to ISR center of mass energies but
it doesn't work above such energy due to the occurrence 
of a shoulder structure in the tail of the distribution, whose 
relevance is growing with energy. This 
fact can be interpreted here as the onset of semi-hard events.
This remark follows from the excellent fits shown by 
the UA5 Collaboration  \cite{Fug} on
final particle MDs in non-single-diffractive events in
$hh$ collisions above 200 GeV   in terms
of the weighted superposition of soft  and semi-hard events, the weight being 
given by the fraction of soft events and the MD of each component being  of
Pascal (negative binomial) type.
The lesson one learns is that "old"  Pa(NB)MD 
regularity is violated at c.m.\ energies higher than ISR when applied 
to  the full sample of events but it continues to be valid when applied
separately at the two  individual components (soft and semi-hard) level.
This result should be compared with the observation that, in \ee\ 
annihilation, the Pa(NB)MD regularity is also 
violated in the full sample of events
(again due to shoulder effect), but it is restored at the  two- and multi-jet 
component level \cite{DEL:4}; 
in addition, residuals analysis on the two-jet sample of 
events indicates   that the regularity in \ee\   should work even better at 
 single quark-jet level \cite{hqlett:3}
(the building block of multiparticle dynamics in
the reaction): accordingly, the
observed final particle multiplicity distribution and the related correlation 
structure are expected to be simply the result of the weighted 
superposition of the mentioned  elementary substructures at different levels 
of investigation.

Finally, KNO scaling is  probably  a correct asymptotic 
(how far?) prediction and  its early  occurrence at lower energies     
should be considered ---as suggested by L\'eon Van Hove---
an inessential transient regime or, as we prefer, limited in $hh$ collisions
to the soft component only. 
This still open question notwithstanding,  we believe KNO 
scaling to be a useful tool in order to describe extreme scenarios of the
production process; in addition it should be remembered that Pa(NB)MD  for
average charged multiplicity $\nbar$ 
larger than $k$  
(with $k>1$) can be written as a  gamma distribution in the scaled variable
$z = n / \nbar$:
\begin{equation}
   \nbar P_n^{\text{(PaNB)}} = \frac{k^k}{\Gamma(k)} z^{k-1} e^{-kz}
\end{equation}
i.e., it has  KNO form. KNO scaling behaviour occurs  
when $k$ becomes an energy
independent parameter.
Remember that the parameter $k$ is linked to the dispersion $D$ by      
\begin{equation}
 k^{-1} = \frac{D^2 - \nbar}{\nbar^2}	\label{eq:k}
\end{equation}
(We notice in passing that, when the
distribution is not a Pa(NB)MD, eq.~\eqref{eq:k} 
can be taken as a definition for
the parameter $k$ in terms of dispersion and average multiplicity).
 
The correlation structure  of multiparticle production in $hh$ 
collisions can be
investigated also by means of other important observables like the factorial 
moments of order $n$ of the multiplicity distributions, $F_n$,
and  the corresponding cumulant moments, $K_n$, 
which are particularly sensitive
to the tail of the distributions and currently used in order to
study intermittent behaviour in very small rapidity intervals.
Of interest is also  the ratio $H_n$  of the above mentioned observables, 
i. e. $K_n/F_n$, when plotted as a function of its order, integer
$n$. It shows 
indeed a characteristic oscillatory behaviour 
already seen in \ee\ annihilation and there explained \cite{hqlett:2}
as due to the same cause which
originated the shoulder effect in the multiplicity distributions in that 
reaction, i.e., the weighted superposition of events of different topology.
It should be pointed out that observed oscillatory behaviour 
of $H_n$ versus $n$   
in $hh$ collisions can also be explained  in terms of the  same
cause which allowed to understand the shoulder effect in final particle MDs,
i.e., the superposition of soft and semi-hard events \cite{mont97}.
These facts, if confirmed, would point in the direction of what is the real 
goal of multiparticle dynamics, an integrated
description of correlation structure and multiplicity distributions in
collisions belonging to the same class, whereas the superposition
mechanism of elementary subprocesses seems to be common to all classes
of collisions.
It is also striking that most elementary substructures in $hh$ collisions as 
well as in \ee\  annihilation, i.e., soft and semi-hard components and
two- and multi-jet contributions  respectively, 
are well described by Pa(NB)MDs.
In this sense Pascal (negative binomial) regularity is still 
valid and  should  be considered even more 
fundamental than originally thought in the field.

 The  aim of  this paper is to explore charged particle multiplicity 
distributions and corresponding  correlation structure  in hadron-hadron 
collisions in the  TeV region in full phase space 
within the  just sketched general 
phenomenological framework by using the above mentioned collective variables.
Accordingly, we propose to describe
MDs  in full phase space  in the new horizon in terms 
of the weighted  superposition of the MDs of soft and semi-hard
events, with each component  assumed to be of Pa(NB)MD type. 
With these two simplifications
the full problem is  reduced  to determine   the energy dependence 
of the Pa(NB)MD parameters, i.e., of  the average charged 
multiplicity, $\nbar$, and 
of parameter $k$
for the soft and semi-hard components. 

It appears that in this essential framework  at least three 
scenarios are possible:
\begin{enumerate}
\item[1.] KNO scaling limit is achieved in the TeV region for both components.
\item[2.] KNO scaling is valid for the soft component, but it is  
heavily violated 
for the semi-hard component in the TeV region.
\item[3.] A QCD-inspired scenario can be obtained by assuming 
that the form of perturbative  
QCD predictions for the width of the MD can be used
also in the non-perturbative sector.
\end{enumerate}

The first two scenarios should be considered quite extreme
possibilities. They determine in a certain sense reasonable bounds to the
variation of the Pa(NB)MD parameters. The third one turns out to give
Pa(NB)MD parameters behaviour intermediate between the first two.

We are aware of the fact that the   assumptions  of the proposed
description are  not unique. They are dictated to us by our experience  in 
the field and by our personal taste. This consideration notwithstanding we 
believe that this research line  should be pursued  in order to give some 
hints to future experimental and theoretical work, and to explore whether  
new phenomena are predicted  by the present approach  at higher energies.
Our findings should be confronted with  results of other possible 
phenomenological  approaches (which we are  demanding) leaving to experiments,
when available, to decide among  different realistic alternatives.

Our aim is to illustrate the Pa(NB)MD composition technique and to apply
it to a simple problem. 
Of course one relevant fraction of events is
expected to come also from diffraction which affects both the soft and
semi-hard components. 
Accordingly, one can study in this context substructures generated by
diffraction in the two previous components by using again the
composition in terms of weighted 
Pa(NB)MDs of diffractive and non-diffractive events.
The use of a Pa(NB)MD for describing diffractive events leads indeed
to results not far from those of a modified gamma distribution
\cite{Goulianos}.
We do not intend to explore for the moment this new perspective, which
we postpone to future work, but which we are ready to use when
a correct residual analysis on MDs will show indications of new
substructures. What is studied in this paper is indeed the structure
of non-single-diffractive events only.

\section{$P_n$ vs $n$ behaviour and $H_q$ vs $q$ oscillations 
in full phase space in the GeV  and in the TeV energy  domains}
%--------------------------------------------------------------

It has been shown in the accelerator 
region that final $n$ charged multiplicity
distributions  in full phase space, $P_n$, are initially
narrower than a Poisson distribution ($1/k$  Pa(NB)MD parameter is negative)
and the distribution is indeed a binomial distribution, produced
particles are very few and like to stay far apart one from the
other, particles anti-correlations seem here to be favored.
Then at approximately 30 GeV c.m.\  energy the observed MD becomes Poissonian:
Pa(NB)MD parameter $1/k$ is zero in this region, particles are produced 
independently one from the other (as predicted by the naive multi-peripheral 
model). Above 30 GeV c.m.\ energies up to ISR energies the distribution is 
a true Pa(NB)MD: $1/k$ parameter becomes positive, the number of produced 
particles 
is larger, two particle correlation are dominant as requested by hierarchical 
correlations structure. A single Pa(NB)MD due to the flexibility of its $1/k$ 
parameter is  describing quite well all the above mentioned 
experimental facts, which are apparently dominated by soft events.
A shoulder structure in the tail starts then to appear at higher energies as 
shown by the  UA5 Collaboration \cite{UA5:3}; 
a single Pa(NB)MD cannot describe the new effect 
(one talks of Pascal (negative binomial) regularity violation), 
which can be interpreted as the onset 
of semi-hard events (events with mini-jets). Notice that Pythia Monte Carlo 
calculations give at present  unsatisfactory results in this area 
\cite{frascati97:AG}, although a careful optimization of the
parameters can of course improve this trend.

Accordingly, as reminded in the introduction, it has been proposed to 
describe  the observed   shoulder structure as the weighted superposition of 
soft events (events without mini-jets) and semi-hard events (events with 
mini-jets), the weight being the fraction of soft events and the MD of each 
component being of Pa(NB)MD type.
The resulting master equation for $P_n$ turns out to be the following
\begin{multline}
   P_n(\asoft; \nsoft, \ksoft; \nsemi, \ksemi) = \\
	\asoft(\rs) P_n^{\text{(PaNB)}}(\nsoft(\rs), \ksoft(\rs)) +\\
	(1 -\asoft(\rs)) P_n^{\text{(PaNB)}}(\nsemi(\rs), \ksemi(\rs))
		\label{eq:combo}
\end{multline}
whose physical content is self-explanatory. 
Notice that we do not
consider interference terms because the classification of events as
soft or semi-hard is based on the final hadronic state, not on the
underlying partonic event.

\begin{figure}
 \begin{center}
  \mbox{\epsfig{file=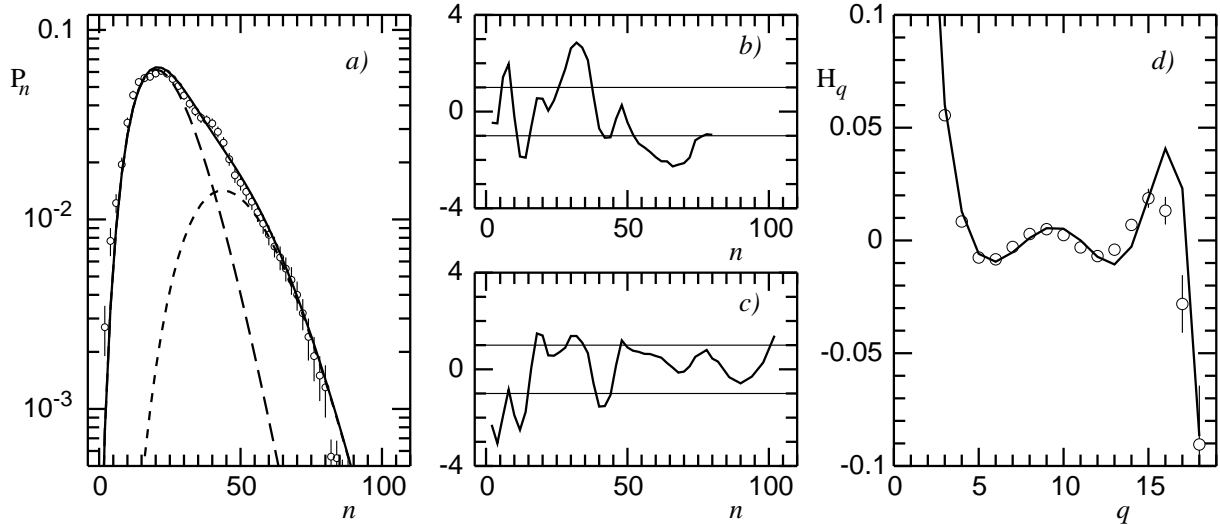,width=\textwidth}}
 \end{center}
 \vspace{-0.5cm}
\caption[Pn, Hq, and residuals at 546 GeV]{\textit{a)} 
Multiplicity distribution at c.m.\ energy 546 GeV  (UA5
data), with the two components of eq.~\eqref{eq:combo},
corresponding residual analysis of \textit{b)}
a fit with one single Pa(NB)MD and of \textit{c)} the fit with
eq.~\eqref{eq:combo}; \textit{d)} ratio of moments  $H_q$,
calculated from eq.~\eqref{eq:combo} after truncation.}\label{fig:PnHq}
\end{figure}

Excellent fits have been obtained for $P_n$ \cite{Fug},
an example of which is shown in Figure~\ref{fig:PnHq}\textit{a}: the plot of
residuals (\textit{b} and \textit{c} in the same figure) shows
how the fit gets better when using two Pa(NB)MDs (although, it should be
mentioned, the fit is not completely satisfactory, which can indicate
the presence of further substructures.
The use of residuals should also be taken \textit{cum grano salis}
in this case, because a minimum chi-square
test has been used to find the Pa(NB)MDs parameters: although the run
test might still be asymptotically independent of the chi-square
test, which does not use information on the sign and sequence of
the deviations, its distribution is in fact not known, thus the
pattern of the residual is only indicative.\footnote{We thank
S.~Krasznovszky for discussion on this point.})

In the fits, the  soft component fraction
decreases from 93 percent at 200 GeV to 72 percent at 900 GeV, $\ksoft$ 
is taken constant as 
the energy increases (as requested by an early KNO scaling behaviour) and its 
best fit numerical value is 7, whereas  $\ksemi$   decreases from 79
at  200 GeV (it describes a nearly Poissonian behaviour) to 13 
at 900 GeV, indicating  strong KNO scaling violation. The average charged
particle multiplicity is approximately two times larger for the semi-hard 
component than for the soft component (as 
observed by UA1 Collaboration \cite{UA1}).
It is interesting (and remarkable) that $H_q$ vs.\ $q$   obtained by data on 
multiplicity distributions oscillates in this region and that the oscillations
are quite  well described by the $K_q$ over $F_q$ ratio calculated  by using 
equation \eqref{eq:combo}, as shown in Figure~\ref{fig:PnHq}\textit{d}.

That's in summary all we know in the GeV region on our variables behaviour.
Coming to the TeV region which we want to explore we will proceed
by extrapolating Equation \eqref{eq:combo} in the new energy domain. 
Although highly simplified our approach still requires
to determine  
the energy dependence of Pa(NB)MD parameters $\nbar$ and $k$ 
for the soft and semi-hard components,
as well as of  soft component fraction $\asoft$, which we propose to
do in the following.

\subsection{$\nsoft$ and $\nsemi$ energy dependence.}
\begin{figure}
\begin{center}\mbox{\epsfig{file=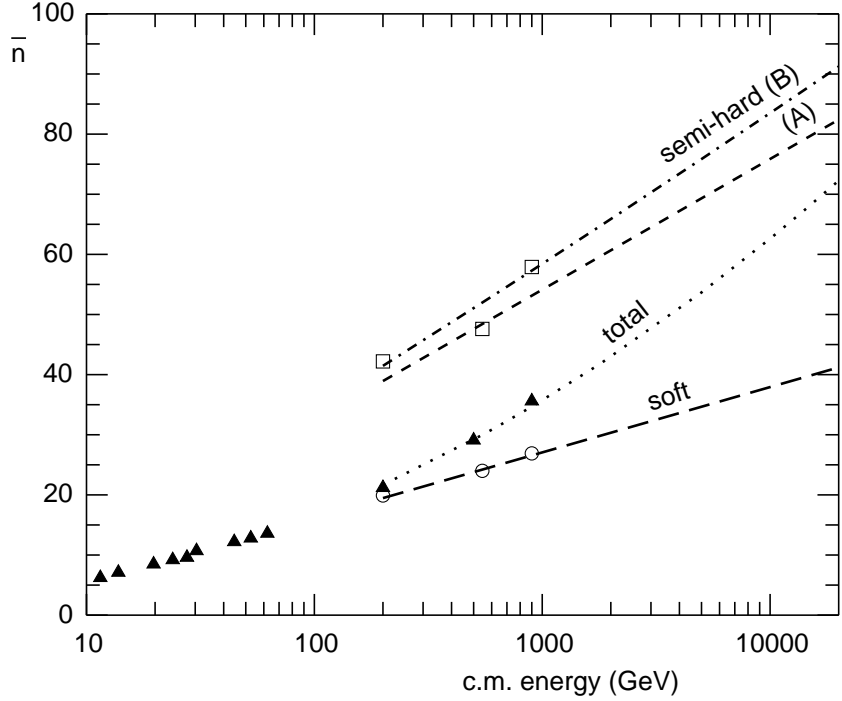}}\end{center}
\caption[average multiplicity]{Average multiplicity $\nbar$ vs.\ c.m.\
energy. The figures shows experimental data (filled triangles) from
ISR and SPS colliders, the UA5 analysis with two Pa(NB)MDs of SPS data
(circles: soft component; squares: semi-hard component), together with
our extrapolations (lines: dotted: total distribution; dashed: soft
component; short-dashed: semi-hard component, eq.\ \eqref{eq:nsemi};
dot-dashed: semi-hard component, eq.\ \eqref{eq:nsemiB}).}\label{fig:nbar}
\end{figure}

%-------------------------------------------------------
For $\nsoft$ it is assumed that its fitted values in the  multiplicity
distributions in the GeV region can be extrapolated to  higher energy domains,
i.e.,
\begin{equation}
\nsoft(\rs) = -5.54 +4.72 \ln(\rs)    \label{eq:nsoft}
\end{equation}
(dashed line in Figure~\ref{fig:nbar}), with $\rs$ in $\text{GeV}$.

\begin{figure}
\begin{center}\mbox{\epsfig{file=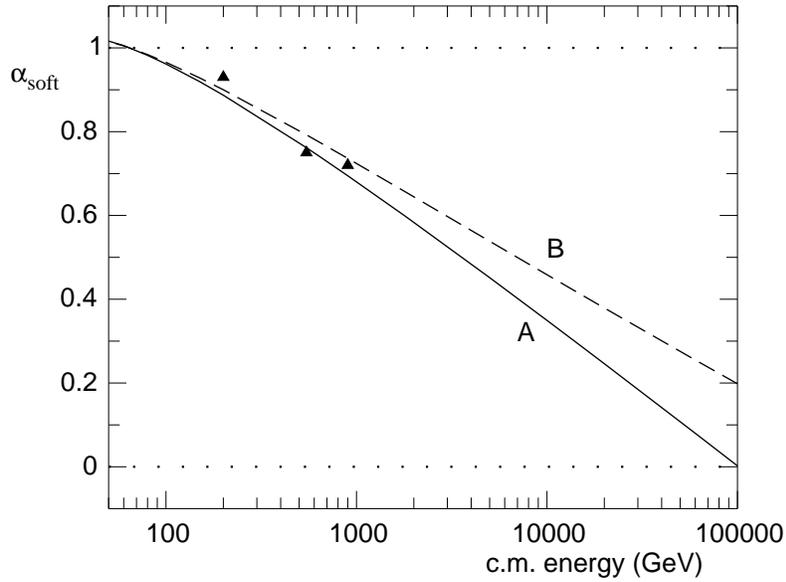,width=10cm}}\end{center}
\caption[energy dependence of alpha soft]{Energy dependence of 
the superposition parameter $\asoft$ (fraction of soft events)
in the two cases of a linear (solid line, eq.\ \eqref{eq:nsemi}) and 
quadratic (dashed line, eq.\ \eqref{eq:nsemiB}) dependence of the
average multiplicity of the semi-hard component on c.m.\ energy.
The triangles are the result of the UA5 analysis \cite{Fug}.
}\label{fig:alphasoft}
\end{figure}

\begin{subequations}
\renewcommand{\theequation}{\theparentequation.\Alph{equation}}
Assuming UA1 analysis on mini-jets  to be  approximately   
valid  also at  higher energies one has  for $\nsemi$
(short-dashed line in Figure~\ref{fig:nbar}):
\begin{equation}
\nsemi(\rs) \approx 2 \nsoft(\rs)        \label{eq:nsemi}
\end{equation}

Alternatively one can postulate that $\nsemi(\rs)$ is 
increasing more rapidly  with energy 
and correct Eq.\ \eqref{eq:nsemi} by adding on  its right side a $\ln^2(\rs)$
term, accordingly one obtains
\begin{equation}
\nsemi(\rs) \approx 2 \nsoft(\rs)   + c' \ln^2(\rs) \label{eq:nsemiB}
\end{equation}
\end{subequations}
(dash-dotted line in Figure~\ref{fig:nbar}).
This simple correction might take into account observed deviations from
Eq.\ \eqref{eq:nsemi} behaviour at 900 GeV and estimate 
from the fit at the  same time parameter $c'$
in Eq.\ \eqref{eq:nsemiB}, which  turns out to be $\approx 0.1$.

Finally  $\ntot$ of the resulting multiplicity distribution  in agreement 
with the common wisdom is given by a quadratic fit (dotted line in 
Figure~\ref{fig:nbar}):
\begin{equation}
\ntot = 3.01 - 0.474 \ln(\rs) +  0.754 \ln^2 (\rs)     \label{eq:ntot}
\end{equation}
                
Being now in this approach  
\begin{equation}
\ntot = \asoft \nsoft + (1 -  \asoft) \nsemi    \label{eq:ncombo}
\end{equation}
the energy dependence of $\asoft$ can easily be determined. It turns out 
to be in the two  cases  described by Eq.s.\ \eqref{eq:nsemi} 
and \eqref{eq:nsemiB}
\begin{subequations}
\renewcommand{\theequation}{\theparentequation.\Alph{equation}}
\begin{equation}
 \asoft = 2 - \ntot/\nsoft      \label{eq:asoft}                %   [6. A]
\end{equation}
and
\begin{equation}
\asoft= 1 + [\nsoft-\ntot]/ [\nsoft  + c'\ln^2(\rs)] \label{eq:asoftB}
\end{equation}
respectively.
\end{subequations}

In Figure \ref{fig:alphasoft} the soft events fraction is shown to be
quickly decreasing with energy; the 
general trend  is to invert the situation observed in the GeV region where 
soft events fraction was dominant: 
semi-hard events fraction is here increasing
from 11 (0.2 TeV) to 75 (20 TeV) percent of the reaction.
Significant changes are introduced by the presence of the $\ln^2 (\rs)$
term in eq.~\eqref{eq:nsemiB} at higher energies: the fraction of
semi-hard events is in this case 10 percent at 0.2 TeV but increases
only to 62 percent at 20 TeV. In the previous case, the semi-hard
component becomes larger than the soft one at 3--4 TeV, but now 
the $\ln^2(\rs)$ term induces this change at approximately
7 TeV. This difference will be visible especially in
scenarios 1 and 3.

\subsection{ $\ksoft$ and  $\ksemi$ energy dependence}
%--------------------------------------------

$\ksoft$ was found  to be constant in the GeV region by the UA5
collaboration; 
since  in addition $\nsoft$ is growing with energy, see Eq.\ 
\eqref{eq:nsoft}, to 
assume $\ksoft$ constant in the new energy domain implies
\begin{equation}
  {D^2}_{\text{soft}}/{\nsoft}^2  
   \approx ~\text{constant}~ \approx 0.143        % [7]
\end{equation}
which corresponds to say that KNO scaling behaviour is valid for the soft
component in the TeV region. We stay with this assumption on $\ksoft$.
It should be pointed out that $\ksoft$ is not affected by the
introduction of the $\ln^2\rs$ term in eq.~\eref{eq:nsemiB}.

As anticipated in the introduction,
the discussion on the  behaviour of $\ksemi$ opens at least three possible
scenarios which are discussed in the following.

\section{The three scenarios}

\subsection{Scenario 1}
\begin{figure}
\begin{center}\mbox{\epsfig{file=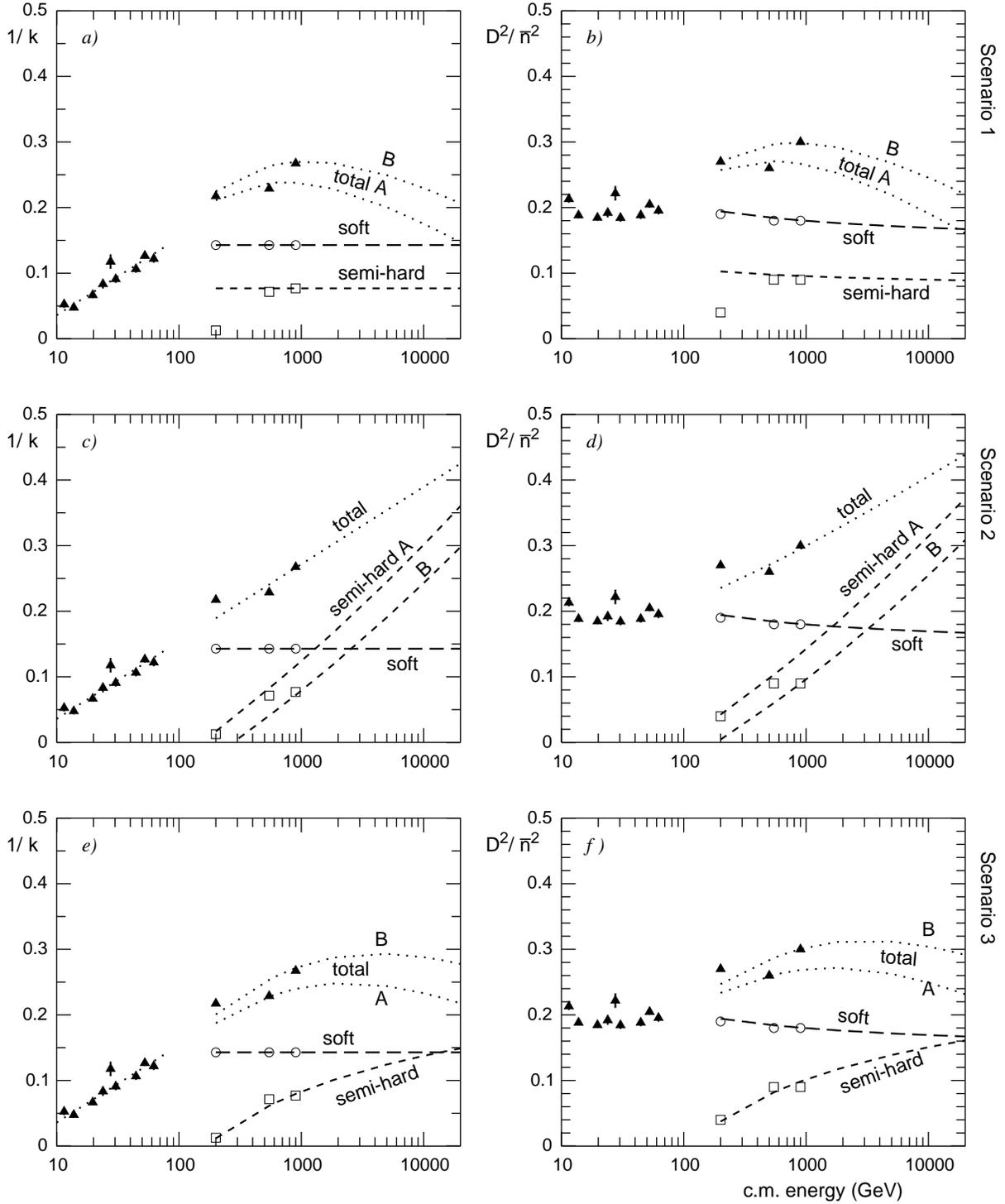,width=\textwidth}}\end{center}
\caption[parameter k and kno variable]{The parameter $1/k$ 
(figures \textit{a,c,e}) and the
KNO scaling parameter $D^2/\nbar^2$ (figures \textit{b,d,f})
are plotted for the scenarios described in the text
(from top to bottom: \textit{a,b}: scenario 1; \textit{c,d}: scenario 2;
\textit{e,f}: scenario 3).
The figures shows experimental data (filled triangles) from
ISR and SPS colliders, the UA5 analysis with two Pa(NB)MDs of SPS data
(circles: soft component; squares: semi-hard component), together with
our extrapolations (lines: dotted: total distribution; dashed: soft
component; short-dashed: semi-hard component).
}\label{fig:k.kno}
\end{figure}

KNO scaling holds in the TeV region also for the semi-hard component, i.e.,
we assume that $\ksemi$ is decreasing until  900 GeV (its value is
13 at this c.m.\  energy) and then it remains  constant in the new region.
Being $\nsemi$ even larger than $\nsoft$,
${D^2}_{\text{semi-hard}} / {\nsemi}^2 \approx 0.09$ throughout 
all the explored energy range. See Figure~\ref{fig:k.kno}\textit{a,b}.

The effect of a quadratic growth of $\nsemi$ with energy, 
eq.~\eqref{eq:nsemiB},
in this scenario is to increase the value of $1/\ktot$ (curves B in
Figure~\ref{fig:k.kno}\textit{a,b}), via the change in $\asoft$,
eq.~\eref{eq:asoftB}. This fact is consistent with our assumption of
the superposition mechanism.

Expected  multiplicity distributions  at 1.8 TeV
and 14 TeV c.m.\ 
energies and corresponding $H_q$ vs.\ $q$ oscillations fitted by using the 
composition of the soft and semi-hard substructures are shown in figure
\ref{fig:MD.scen1}. The energies of 1.8 and 14 TeV have been chosen
because they are respectively the c.m.\ energy of \Tevatron\ and the
expected c.m.\ energy of LHC.
It is interesting to remark that the shapes of the multiplicity distributions
of the two components are similar at all energies, but the heights of the
corresponding maxima are reversed in going from the lowest to the top energy.
In addition oscillations seem to be stretched in shape as  the energy
increases, indicating their tendency to be  highly reduced at higher
energies (notice that we show here $H_q$ moments calculated without
truncating the MD, because the truncation depends solely on the
size of the experimental sample; thus we only show the moments
computed for the total MD, as the two components are each of Pa(NB)MD type and
therefore, individually considered, show no oscillations).

A rather large change in the shape of the MD is introduced when we
consider the quadratic term for $\nsemi$, eq.~\eqref{eq:nsemiB}:
even at 20 TeV the maximum of the semi-hard component has not yet become
larger than that of the soft component. On the other hand, the
shoulder has become more evident, due to the higher average
multiplicity that was introduced for the semi-hard component
and the smaller resulting value of $\asoft$;
at a c.m.\ energy of 14 TeV the area of the maximum of the total MD is
rather flat and shows a small dip (see fig.~\ref{fig:MD.scen1}\textit{a}). 
Correspondingly the oscillations of
the $H_q$ moments become approximately 4 times larger in amplitude,
but don't vary much in period, and the
first minimum is not shifted.

\subsection{Scenario 2.}
The main assumption is that the general trend observed in the GeV region
for the total distribution continues to be valid in the TeV region, i.e.,
$D^2_{\text{total}} / \ntot^2$ is logarithmically growing, suggesting 
strong KNO scaling behaviour violation:
\begin{equation}
  \ktot^{-1} = -0.082 + 0.0512 \ln \rs
\end{equation}
See Figure~\ref{fig:k.kno}\textit{c,d}.

The effect of a quadratic growth of $\nsemi$, eq.~\eqref{eq:nsemiB},
in this scenario is to decrease the value of $1/\ksemi$ (curves
labelled  B
in the figure). This fact is again consistent with $1/\ktot$ growing
with energy.

In Figure~\ref{fig:MD.scen2} are shown the multiplicity distributions and the
corresponding $H_q$ vs. $q$ oscillations of this scenario.

The first remark is that the shapes of the two components are totally
different, a wide  queue in the semi-hard component is visible in the high
multiplicity channels. The heights at the maxima favor initially 
the soft component and then the heights of the two components are almost 
equal. It is  striking that  $H_q$ vs.\ $q$ 
oscillations disappear as the c.m.\ 
energy increases,  indicating that 
single Pa(NB)MD behaviour is the dominant feature
in this energy domain.

In this case the effect of a quadratic growth of $\nsemi$ is much less
noticeable than in scenario 1 
in the shape of the MD: the shoulder is only slightly more visible
and the amplitude of the $H_q$
oscillations is moderately larger than in the case of 
linear $\nsemi$.

\subsection{Scenario 3.}
This is the QCD inspired scenario. QCD predicts, 
at the leading  order\footnote{in view of the approximations
involved, more sofisticated calculations \cite{Dokshitzer} are not useful in
this framework.}, for
the parameter $k$ of the multipicity distribution
\cite{Nason}
\begin{equation}
 k^{-1} = a + b \sqrt{\alpha_{\text{strong}}}
		\label{eq:ksemi3}
\end{equation} 
where
\begin{equation}
  \alpha_{\text{strong}}  \approx 1/\ln(Q/Q_0) 
\end{equation}
and $Q$, $Q_0$ are the initial virtuality and the cut-off of the 
parton shower. QCD predicts (for \ee\ annihilation) $a \approx 0.33$
and $b \approx -0.9$, but in order to apply the above equation to our
problem we leave the $a$, $b$ and $Q_0$ parameters free.

Since the constants can be determined  by a least square fit to the 
values found for $k$ at c.m.\  energies 200 GeV, 500 GeV and 900 GeV 
one obtains, assuming Eq.\ \eqref{eq:ksemi3} 
to control $\ksemi$ component behaviour,
\begin{equation}
 \ksemi^{-1} = 0.38 - 0.42 / \sqrt{\ln(\rs/10)}
\end{equation}
The result is presented in Figure~\ref{fig:k.kno}\textit{e,f}. 
Notice how $1/\ktot$
is indeed lower than in scenario 2 but higher that in scenario 1.

The effect of a quadratic growth of $\nsemi$, eq.~\eqref{eq:nsemiB},
in this scenario is once again to increase the value of $1/\ktot$
similarly to what happens in scenario 1.

\def\labelfigPnHq#1{
   \put(5.7,20.2){\scriptsize \textsc{Scenario #1.A}}
   \put(12.1,20.2){\scriptsize  \textsc{Scenario #1.A}}
   \put(5.7,14.65){\scriptsize  \textsc{Scenario #1.B}}
   \put(12.1,14.65){\scriptsize  \textsc{Scenario #1.B}}
   \put(5.7,8.7){\scriptsize  \textsc{Scenario #1.A}}
   \put(13.1,8.7){\scriptsize  \textsc{Scenario #1.A}}
   \put(5.7,3.7){\scriptsize  \textsc{Scenario #1.B}}
   \put(13.1,3.7){\scriptsize  \textsc{Scenario #1.B}}
}
\begin{figure}
\begin{center}
\mbox{\epsfig{file=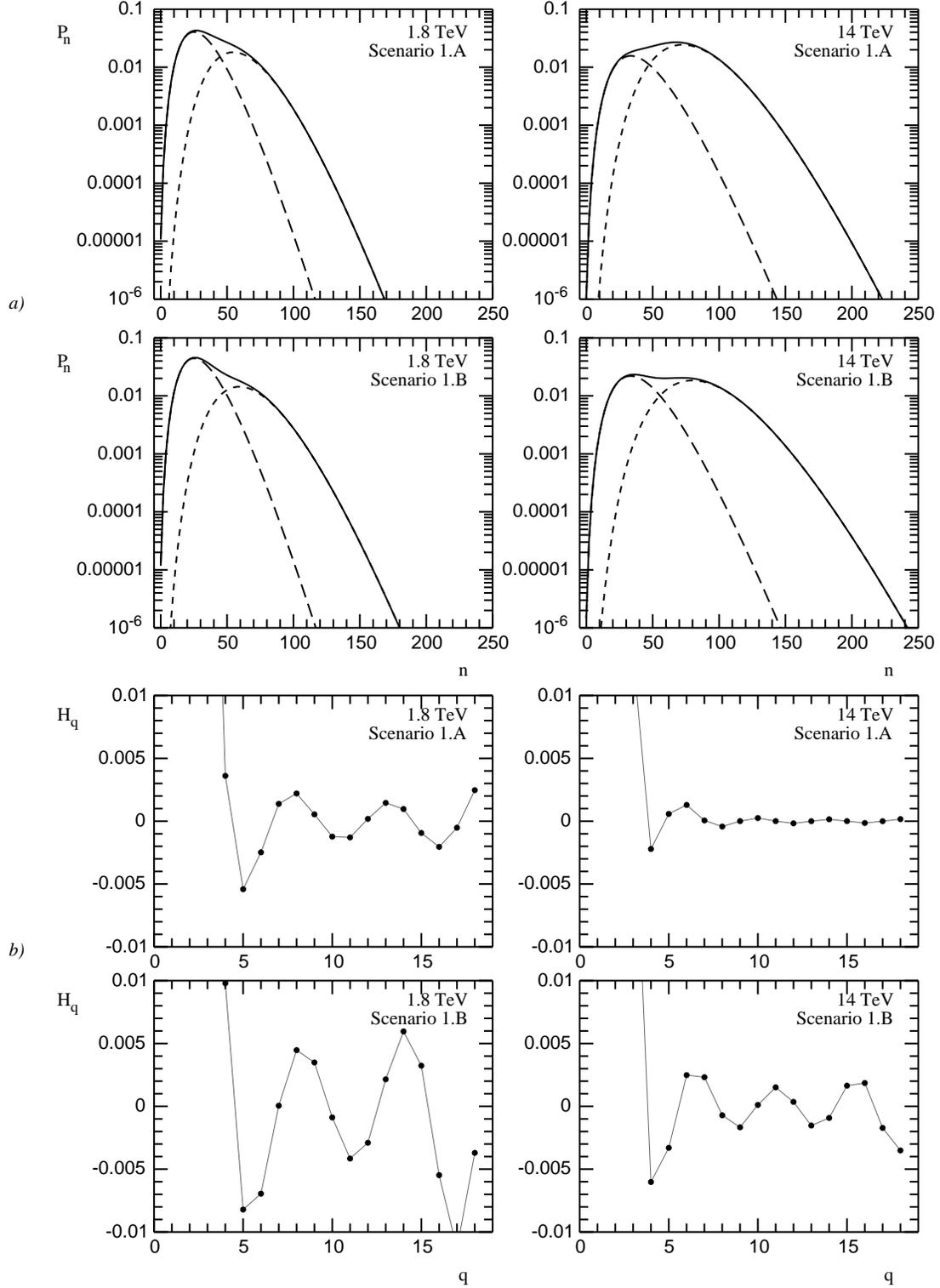,height=20cm}}
\end{center}
\caption[MD and Hq - scenario 1]{\textit{a)} 
Multiplicity distributions for scenario 1 at c.m.\  energies 
of 1.8 (\Tevatron\ energy) and 14 (LHC expected energy) TeV;
the first row refers to solution A for the average multiplicity
in semi-hard events, the second row to solution B
(solid line: total
distribution; dashed line: soft component; short-dashed line:
semi-hard component). \break
\textit{b)} filled circles:
$H_q$ vs $q$ oscillations (without truncation of the MD) 
corresponding to the total distribution of part \textit{a};
the line is drawn to guide the eye; again
the first row refers to solution A for the average multiplicity
in semi-hard events, the second row to solution B.
}\label{fig:MD.scen1}
\end{figure}

\begin{figure}
\begin{center}
\mbox{\epsfig{file=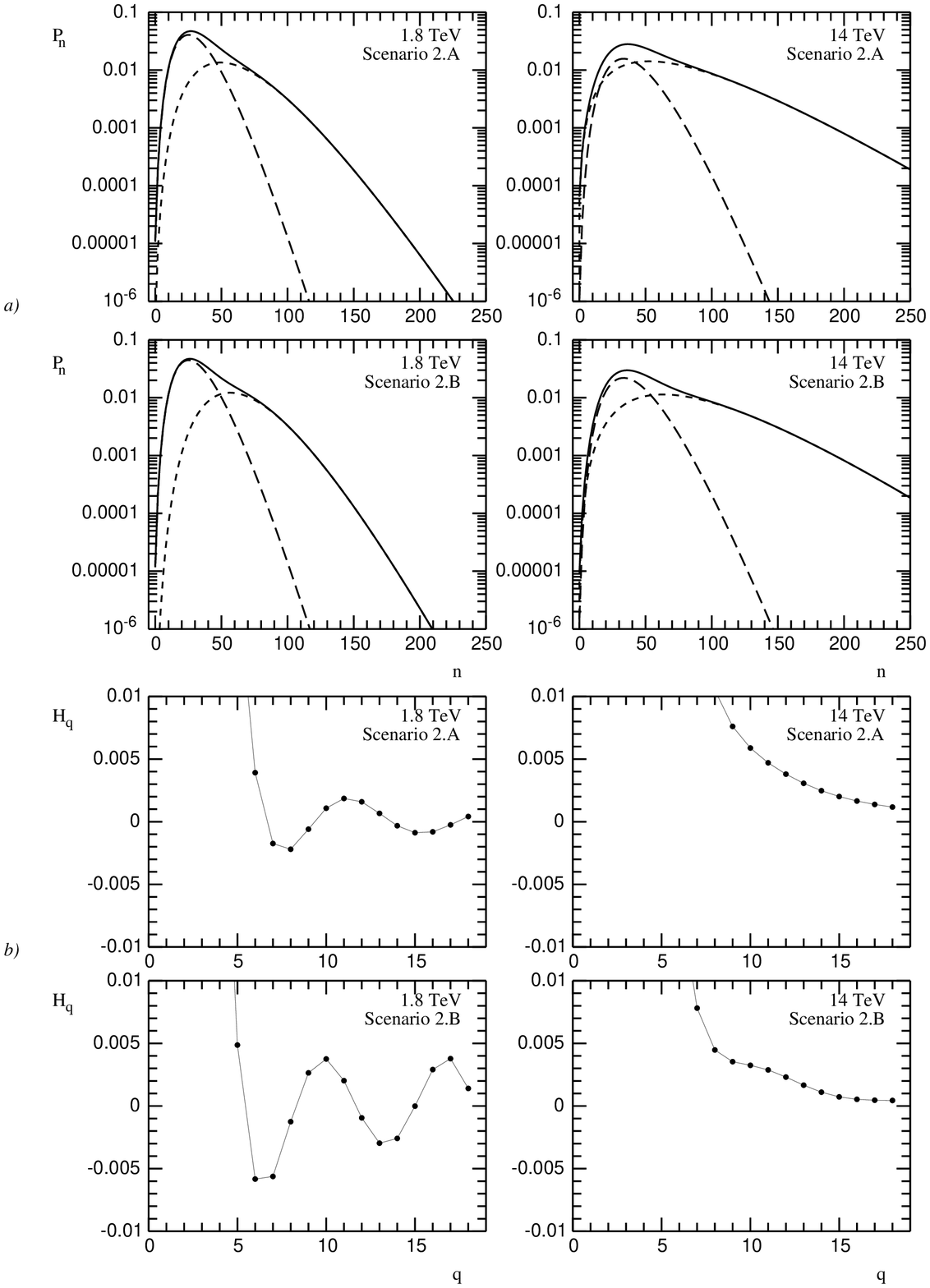,height=20cm}}
\end{center}
\caption[MD and Hq - scenario 2]{Same content 
as Figure~\ref{fig:MD.scen1} but for scenario 2.
}\label{fig:MD.scen2}
\end{figure}

\begin{figure}
\begin{center}
\mbox{\epsfig{file=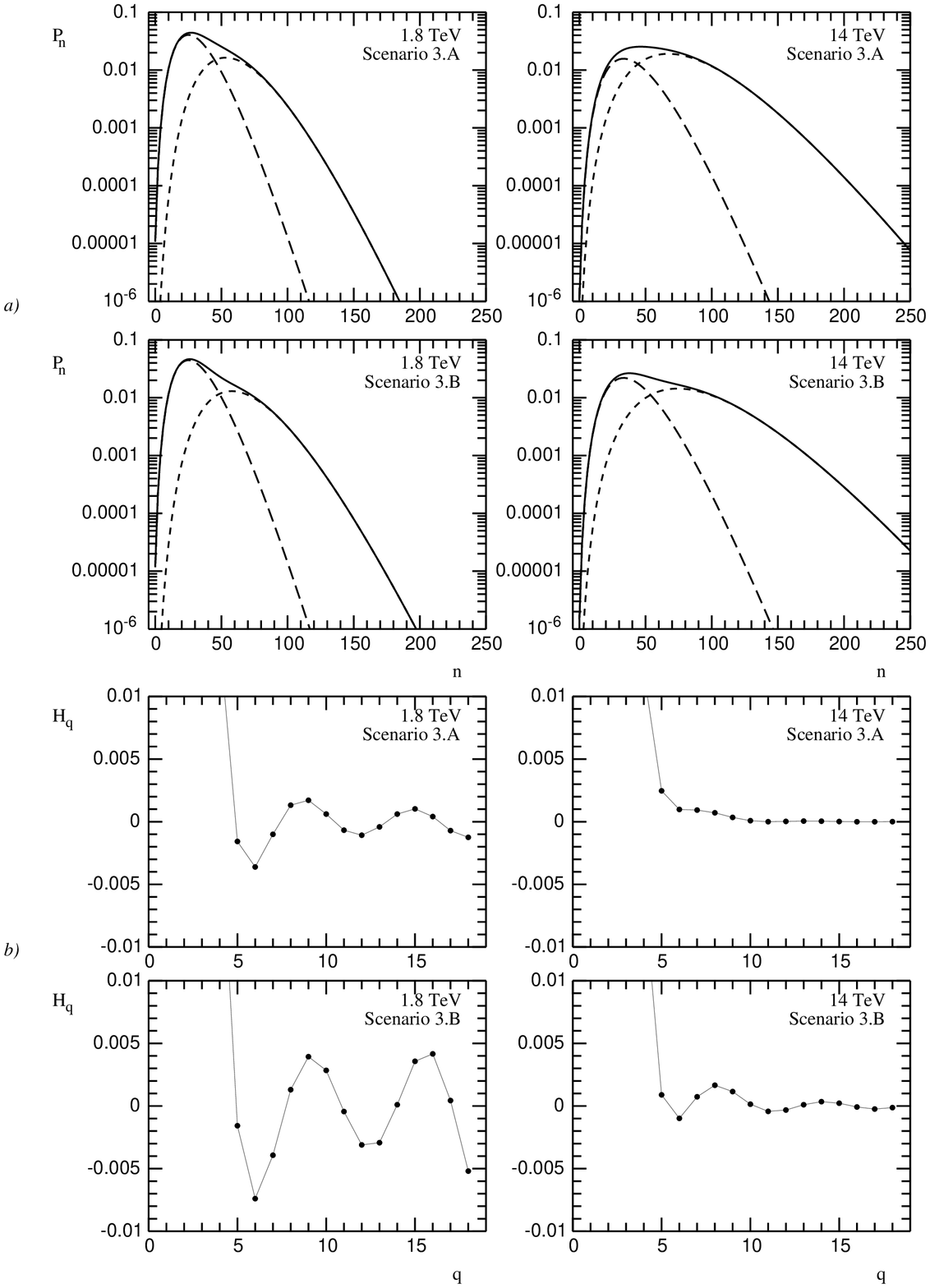,height=20cm}}
\end{center}
\caption[MD and Hq - scenario 3]{Same content 
as Figure~\ref{fig:MD.scen1} but for scenario 3.
}\label{fig:MD.scen3}
\end{figure}

The new situation for $P_n$ vs $n$ and $H_q$ vs $q$
is summarized in Fig.~\ref{fig:MD.scen3}.
It is interesting to remark that this scenario gives  predictions
on both variables which
are  intermediate between the two previous extreme ones of scenarios 1 and 2.
In Figure \ref{fig:MD.scen3}\textit{a} 
one sees in fact  that the tail of  $P_n$ vs $n$
is increasing with c.m.\  energy  but high multiplicity channels are not
populated as in Figure~\ref{fig:MD.scen2}, 
although they are larger than those in Figure~\ref{fig:MD.scen1}.
Mini-jets production is intermediate between the two scenarios.
Accordingly, in Figure \ref{fig:MD.scen3}\textit{b} 
$H_q$ vs $q$ oscillations are decreasing with c.m.\ 
energies but not as much as in scenario 2, indicating that the absence of
oscillations is here an asymptotic prediction  and coincides with 
expectations of  $H_q$ oscillations  description  in terms of a single
Pa(NB)MD. It is in fact quite clear that in the limit $\asoft \to 0$
mini-jets production is dominant
with respect to soft events and the corresponding multiplicity distribution 
is described almost fully  by a single Pa(NB)MD.

The effect of a quadratic growth of $\nsemi$ is quite noticeable here,
as the shoulder structure, almost disappeared above 10 TeV with linear
$\nsemi$ is now well visible up to 20 TeV, because the semi-hard
component has not yet become dominant. Correspondingly, $H_q$
oscillations are approximately 3 times as large as in the linear case.

\bigskip
It appears that  probably it is   a too ``black and white'' attitude to 
assume for $\ksemi$ strong KNO scaling  behaviour (scenario 1) or strong 
KNO scaling violation (scenario 2).
These too extreme choices were done of course on purpose 
in order to fix  the boundary conditions to our exploration: the real world 
might very well be (as illustrated by scenario 3)  
between the above two.

In conclusion the reaction is controlled by the ratio of soft to semi-hard
events. This ratio favors soft events production up to ISR energies
and a single Pa(NB)MD describes here quite well all experimental facts, above 
such energies semi-hard events start to play a more important role and are 
revealed by two effects
(the onset of shoulder structure in the multiplicity distributions and 
$H_q$ vs $q$ oscillations in  related  correlations); both effects can
be cured by using the weighted 
composition of two Pa(NB)MDs, one for the soft part of
the  reaction, and  the second for its semi-hard part.
In the TeV region semi-hard events become dominant, they obscure soft events
production and following our assumptions asymptotically ($\asoft \to 0$)
a single Pa(NB)MD is describing again quite  well 
multiplicity distributions and corresponding correlation structure.

\section{Clan  analysis of the  soft and semi-hard component
substructures}
%-------------------------------------------------------------------

\begin{figure}
\begin{center}
\mbox{\epsfig{file=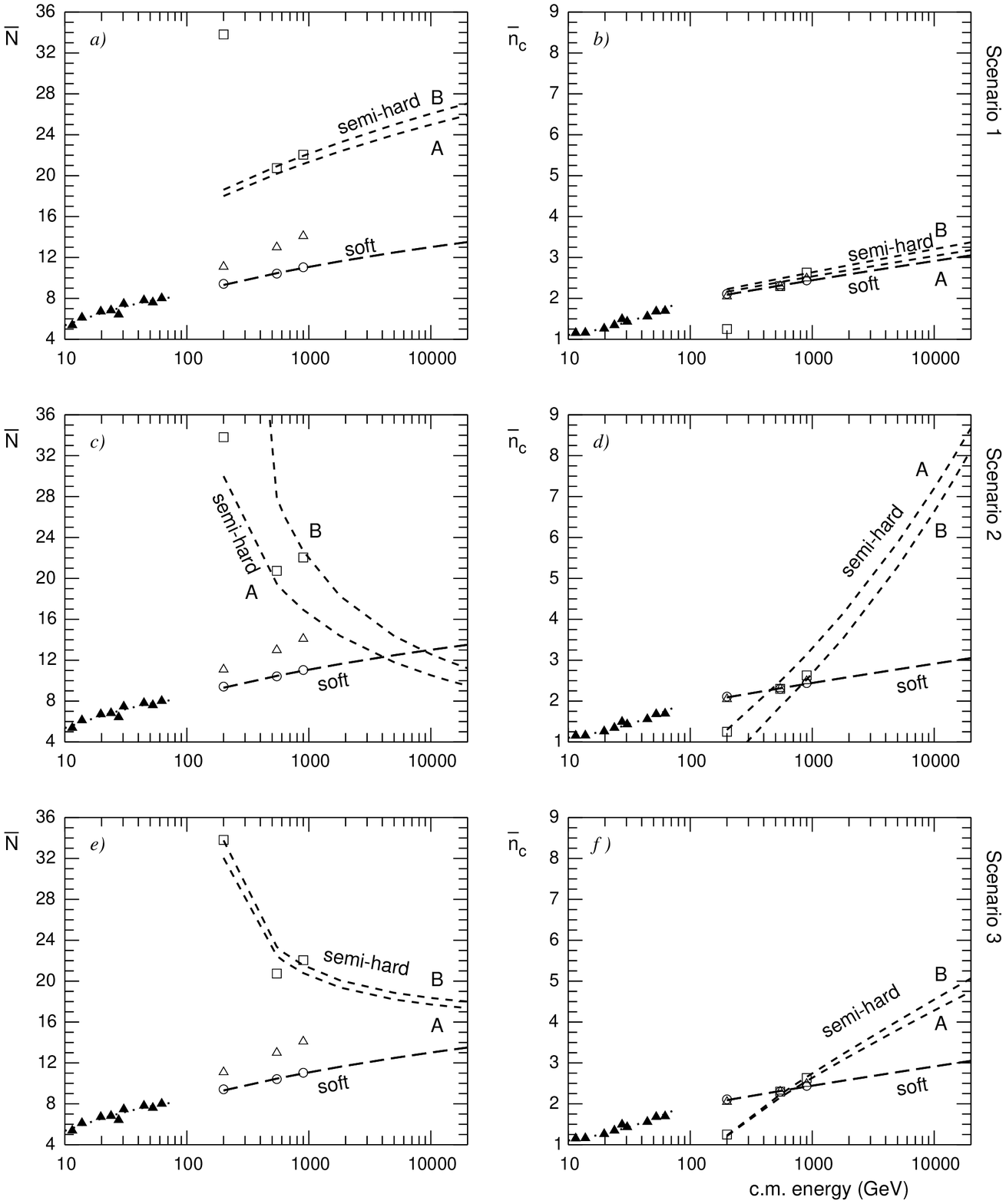,width=\textwidth}}
\end{center}
\caption[Clan parameters]{Clan parameters $\Nbar$
(figures \textit{a,c,e}) and $\nc$ (figures \textit{b,d,f})
are plotted for the scenarios described in the text
(from top to bottom: \textit{a,b}: scenario 1; \textit{c,d}: scenario 2;
\textit{e,f}: scenario 3).
The figures shows experimental data (filled triangles) from
ISR and SPS colliders, the UA5 analysis with two Pa(NB)MDs of SPS data
(circles: soft component; squares: semi-hard component), together with
our extrapolations (lines: dotted: total distribution; dashed: soft
component; short-dashed: semi-hard component).
}\label{fig:clans}
\end{figure}

The decision taken by UA5 Collaboration  to  describe  soft and semi-hard 
events  component  substructures in full phase space in the GeV region in 
terms of Pa(NB)MDs and here extended to the TeV region allows to interpret 
all the  above mentioned results in the framework of clan  
analysis \cite{AGLVH:1,FaroAG},
that is we express and comment now our previous results in terms of the two
parameters:
\begin{equation}
  \Nbar = k \ln\left( 1+\nbar/k \right) \qquad\qquad 
  \nc = \nbar / \Nbar			\label{eq:clandef}
\end{equation}
See Figure~\ref{fig:clans}.
The above definition is valid for a single Pa(NB)MD.
Notice that for the total multiplicity distribution, clans cannot be defined:
the fact, as shown in Figures~\ref{fig:MD.scen1}--\ref{fig:MD.scen3}, that
the total MD presents oscillations in the ratio of moments, $H_q$,
implies, via the theorems proven in \cite{hqlett}, that the total MD
is not an infinitely divisible distribution (IDD): indeed only for
IDDs it is possible to generalize the definition of clans 
from that of eq.~\eref{eq:clandef} to that of
intermediate sources produced according to a Poisson distribution.
This is the reason why we will discuss the behaviour of
clan parameters only for each component separately.

The soft component substructure is the same  in all three scenarios. The 
average number of clans is  here a slowly increasing variable with
c.m.\ energy
and the 
average number of particles per clan goes from 2.6 at 1.8 TeV to 
3.1 at 20 TeV, indicating that phase space is  homogeneously filled by
independent almost equal size clans  as   requested by a KNO scaling regime.
One can talk in this context pictorially  of independent equally populated
sources whose number is a slowly increasing function of 
available  c.m.\  energy.

In scenario 1, which assumes KNO scaling behaviour also for the semi-hard
component, the average number of particles per clan is approximately the same
as  that seen in the soft component substructure, whereas the average number
of clans is two times larger in the semi-hard component than in the 
soft one. This finding is consistent with the assumption of equation
\eqref{eq:nsemi}.

Scenario 2 shows  a dramatic increasing  with energy of the 
$D^2_{\text{semi-hard}}/\nsemi^2$ ratio 
as requested by   strong KNO scaling violation.
Accordingly  the average number of clans is a quickly decreasing 
function of energy 
and then
it becomes  a very slowly decreasing quantity. %(it is 9.5 at 20 TeV)
This behaviour should be confronted with that of the average number of 
particles per clan which at 5 TeV is almost two times larger than at 900 GeV 
and becomes three times larger at 20 TeV. %($ \bar n_{c,\text{semi-hard}} 
Notice that this  huge cascading phenomenon is associated with a limited 
number  of  independent intermediate sources and this is striking.
The two scenarios are indeed ---as already mentioned--- quite extreme.
The highly ordered and homogeneous structure of phase space in scenario 1  
becomes here highly inhomogeneous favoring huge branching production in each 
source to be compared with the fully independent production of the  sources 
(clans). Available c.m.\ energy goes more in particle production within a clan
than in clan production, contrary to what was found in scenario  1.

Finally scenario 3 which it should be remembered is a QCD inspired scenario.
KNO scaling violation for the semi-hard component
although effective  is not as strong as in scenario 2. The 
average number of clans is 18.2 at 5 TeV  (about the same size
seen at 900 GeV) and 17.5 at 20 TeV, suggesting an
almost energy independent production  of the average number of clans over a 
large fraction of the TeV region; in addition the average number of particles 
per clan is  growing with energy but not as much as in scenario 2,
suggesting a moderate average branching consistent with particle production 
in mini-jets.

In Figure~\ref{fig:clans} $\Nbar$ and $\nc$ are plotted  for the individual 
substructures of the collision in the three  scenarios as functions of
the  c.m.\  energy.

It is interesting to remark here a linear growth of $\Nbar$ with the
maximum allowed rapidity for the soft component 
and for the semi-hard component
of scenario 1 to be contrasted with an almost parabolic decrease of
the average number of clans for the semi-hard component in scenarios 2 and 3 
with respect to the same variable.
It is  also remarkable that the average number of clans
is more rapidly decreasing in scenario 2 than in scenario 3,
indicating the occurrence of more branching in each clan.
In addition in scenario 3 the starting of a very slowly decreasing
region for $\Nbar$
is already clearly  visible at 10 TeV  and expected to develop  over a large 
sector of the TeV region.

The effect of a quadratic growth of $\nsemi$, eq.~\eqref{eq:nsemiB},
is visible only in scenario 2, being negligible in scenarios 1 and 3;
obviously the effect appears only in the semi-hard component, and not
in the soft one. 
These facts are a consequence of how we constructed our scenarios:
indeed in scenarios 1 and 3 the term $\ln^2(\rs)$ appears only in $\nbar$
inside the logarithm (see eq. \eqref{eq:clandef}), 
whereas in scenario 2 it appears also in $\ksemi$.
The average number of clans is increased by about 20 percent, while the
average number of particles per clan is decreased by approximately 10
percent.

It should be pointed out that the mild energy dependence of the average number
of clans in full phase space (f.p.s.) over a given energy interval for a 
(soft and/or semi-hard events) substructure of Pa(NB)MD type has important 
consequences:

1) the probability  to produce  empty events (events without charged 
particles) is also mildly energy dependent in view of the equation
\begin{equation}
 P_0 = \exp (-\Nbar)
\end{equation}

2) this fact establishes  a upper bound  to the production of clans 
at any energy $\Nbar(\delta y) \leq \Nbar(f.p.s.)$ and a 
lower bound  to the production of events without
particles in any rapidity interval at a given c.m.\  energy  
$P_0(f.p.s.) \leq P_0(\delta y)$.

\section{New experimental data}
\begin{figure}
\begin{center}\mbox{\epsfig{file=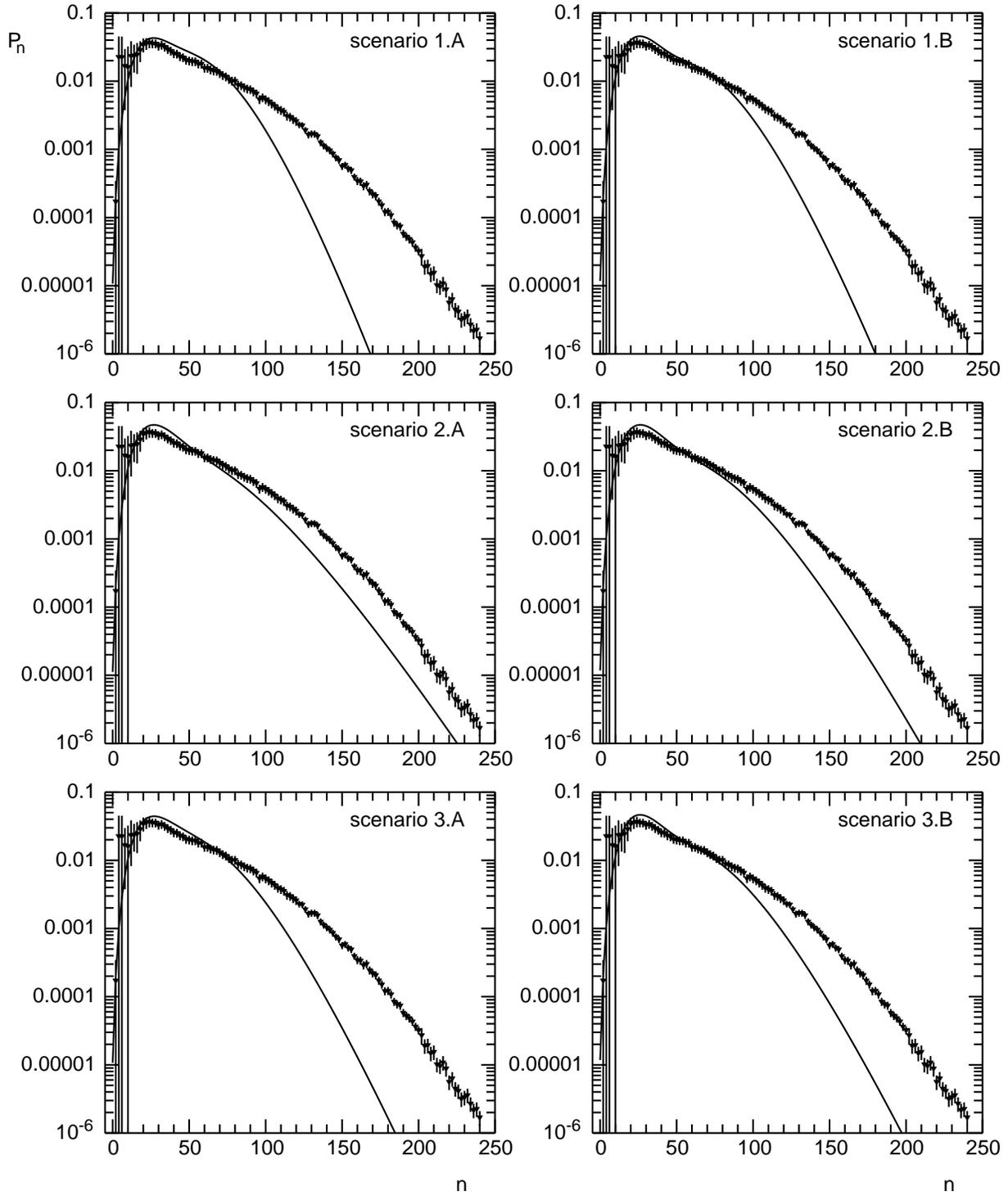,width=\textwidth}}\end{center}
\caption[e735]{Comparison
of our predictions on MDs at 1.8 TeV c.m.\ energy (solid lines)
with the recent data published by the E735 Collaboration (triangles)
\cite{Walker} for each scenario and each option.}\label{fig:e735}
\end{figure}

After we finished our work, the paper by Matinyan and Walker
\cite{Walker}
appeared, providing data from the E735 experiment at \textsc{Fermilab}.
The Authors of that paper find that MD data in f.p.s. up to 1800 GeV
c.m.\ energy can be described in terms of events belonging to two
independent classes, only one of which obeys KNO scaling.
The threshold for the appearance of the second class of events is
estimated approximately between 100 and 200 GeV.
These results agree with the starting point of the present analysis.
The Authors of \cite{Walker} interpret the two classes as events
generated by one collision and by more than one collision, respectively.
Alternatively, we prefer to interpret the two classes as soft and
semi-hard events.
It should be remarked that in the data presented by the E735
experiment the MD are systematically wider than the corresponding
ones of the UA5 Collaboration.
This consideration notwithstanding, we decided to compare the new data
at 1.8 TeV c.m.\ energy to our predictions, presented in the previous
sections. 
As can be seen in Fig.~\ref{fig:e735}, just by inspection, scenario 2
with option A seems to be favored.\footnote{
A more detailed discussion on the experimental results of the E735
Collaboration and their relevance for the study of MDs in the TeV
region is postponed to a forthcoming letter.}
We conclude that according to our description one should expect strong
KNO scaling violation in the MD and logarithmic growth with energy of
the average charged multiplicity of the semi-hard component.
It is interesting to remark that what we considered an extreme
situation turns out to be less extreme from an experimental point of
view:
it is clear indeed that a huge mini-jet production will be the main
characteristic of the new energy domain.
Assuming that observed deviations of E735 results
from our predictions of scenario 2
in fig.~\ref{fig:e735} will be confirmed, they imply that
our $\ksemi$ parameter decreases more rapidly than in our scenario.
This fact has important consequences since the mentioned deviations
occur mainly in the tail region of the distribution, where the
multiplicity is higher.
Accordingly, the integrated two-particle correlation,
$\int C_{2,\text{semi-hard}}(\eta_1,\eta_2) d\eta_1 d\eta_2 = 
\nsemi^2/\ksemi$, are much larger.
Therefore one should expect here a production of  more densely populated
mini-jets characterized by a higher internal two-particle correlation
structure.
Whether this is the onset of a new component to be added to the
previous ones or not will be decided by future experiments at LHC.

\section{Summary}
Possible scenarios of multiparticle production in hadron-hadron collisions
in the TeV region  have been discussed. It has been shown that the most 
spectacular facts are here the occurrence of two classes of events and in 
this framework the dominance of  semi-hard events with respect to the soft 
ones;
this last result should be confronted with the behaviour of the two components
in the GeV region where  just the opposite occurs, i.e. the  soft component
is dominant. Assuming that soft component events multiplicity distributions 
behave according  to KNO scaling expectations,
two extreme situations  for the semi-hard component structure  of the
multiplicity distributions, i.e. an effective KNO scaling regime and a strong 
KNO scaling violation regime, have been compared to a QCD inspired set
of predictions. Essential ingredient of the analysis have been to think to the
final charged particle multiplicity distributions at various energies in terms
of the weighted superposition  of the two above mentioned  components, the 
weight being the fraction of soft events. In addition  following
the success of the fits in the GeV region each component has been assumed to 
be of Pa(NB)MD type.
Scenario 1 is apparently excluded by E735 data at 1.8 TeV c.m.\ energy and
scenario 2, when compared with the same set of data, turns out
to be less extreme than previously thought suggesting that the main feature
in the new region, in our framework, is a huge mini-jet production
together with a possible production of a new species of mini-jets.
The QCD inspired scenario amazingly  gives predictions which
are  intermediate between the other two.

Clan structure analysis when applied to the identified substructures
of the reaction in full phase space in the TeV region reveals unexpected
and interesting  properties which might be relevant for studying
empty events production and bounds to rapidity gaps.

\section*{Acknowledgments}
We would like to thank Prof.~W.D.~Walker for useful discussions on his work.

\section*{References}
%auto-ignore
%%%% References file: substruct-fps.ref
%%%% Creator: REFLATEX.PL 1.0
%%%% Usage: \input{substruct-fps.ref} where references should go.
%%%%

\end{document}